\documentclass[pra,twocolumn,showpacs,superscriptaddress,preprintnumbers,amsmath,amssymb,aps]{revtex4-1}
\usepackage{mathrsfs}
\usepackage{amsmath}
\usepackage{amssymb}
\usepackage{graphicx}
\usepackage{dcolumn}
\usepackage{bm}
\usepackage{subfigure}
\usepackage{color}
\usepackage{ifpdf}
\usepackage{epstopdf}
\usepackage{CJK}
\usepackage{verbatim}
\usepackage{bbm}

\usepackage[
pdfstartview=FitH,%
bookmarks=true,%
bookmarksopen=true,%
breaklinks=true,%
colorlinks=true,%
linkcolor=blue,anchorcolor=blue,%
citecolor=blue,filecolor=blue,%
menucolor=blue,pagecolor=blue,%
urlcolor=blue]{hyperref}
\hypersetup{hypertex=true,
	colorlinks=true,
	linkcolor=blue,
	anchorcolor=blue,
	citecolor=blue}

\usepackage{float}
\usepackage[toc,page,title,titletoc,header]{appendix}

\def\be{\begin{equation}}
\def\ee{\end{equation}}
\def\bea{\begin{eqnarray}}
\def\eea{\end{eqnarray}}
\def\nn{\nonumber}

\begin{document}
\begin{CJK*}{GBK}{song}

\title{Chiral edge states in position shaken finite-size honeycomb optical lattice}
\author{Zhongcheng Yu}
\affiliation{State Key Laboratory of Advanced Optical Communication Systems and Networks, Department of Electronics, Peking University, Beijing 100871, China}
\author{Jinyuan Tian}
\affiliation{State Key Laboratory of Advanced Optical Communication Systems and Networks, Department of Electronics, Peking University, Beijing 100871, China}
\author{Fansu Wei}
\affiliation{State Key Laboratory of Advanced Optical Communication Systems and Networks, Department of Electronics, Peking University, Beijing 100871, China}
\author{Xuzong Chen}
\affiliation{State Key Laboratory of Advanced Optical Communication Systems and Networks, Department of Electronics, Peking University, Beijing 100871, China}
\author{Xiaoji Zhou}\email{xjzhou@pku.edu.cn}
\affiliation{State Key Laboratory of Advanced Optical Communication Systems and Networks, Department of Electronics, Peking University, Beijing 100871, China}
\affiliation{Collaborative Innovation Center of Extreme Optics, Shanxi University, Taiyuan, Shanxi 030006, China}
\date{\today}

\begin{abstract}
The quantum anomalies at the edges correspond to the topological phases in the system, and the chiral edge states can reflect bulk bands' topological properties. In this paper, we demonstrate a simulation of Floquet system's chiral edge states in position shaken finite-size honeycomb optical lattice. Through the periodical shaking, we break the time reversal symmetry of the system, and get the topological non-trivial states with non-zero Chen number. At the topological non-trivial area, we find chiral edge states on different sides of the lattice, and the locations of chiral edge states change with the topological phase. 
Further, gapless boundary excitations are found to appear at the topological phase transition points. It provides a new scheme to simulate chiral edge states in the Floquet system, and promotes the study of gapless boundary excitations.
\end{abstract}

%\pacs{32.80.Qk,37.10.Jk,02.30.Yy,03.75.-b}

\maketitle
\end{CJK*}

\section{Introduction}
Since the quantum Hall effect was discovered \cite{PhysRevLett.49.405,PhysRevLett.50.1953}, the research on topological phases has attracted intense attention. With the study on those topological phenomena developing, edge states are frequently found to be associated with topological properties of bulk bands \cite{PhysRevB.27.5142}, and this relationship is summarized as bulk-edge correspondence \cite{PhysRevB.48.11851,PhysRevLett.71.3697,PhysRevB.74.045125}. Due to topological protection, edge states are robust against weak disorders \cite{Gao2016,PhysRevLett.119.246402,PhysRevB.98.165148}. Consequently there is considerable potential in the design of dissipationless or low-power electronic devices \cite{Ren_2016}, as well as in spectroscopic techniques \cite{Hsieh919,Roushan2009,PhysRevLett.105.076801}. One interesting edge phenomenon is gapless boundary excitation \cite{PhysRevB.43.11025,PhysRevB.84.235141}. For example, graphene exhibits edge states under some particular boundaries, which is of great significance to electronic transport \cite{PANTALEON2018191}. 
There are several detection methods of edge states topological materials \cite{Drozdov2014}: The most commonly used method is detecting quantized values on semiconducting heterostructures \cite{Roth294,PhysRevLett.107.136603}, and some unique topological semimetals, like Weyl semimetals, can be directly observed by fermi arcs \cite{Xu613,Inoue1184,Batabyale1600709,Drozdov2014}.

Recently, many researchers have begun to explore topological phenomena in Floquet systems, a kind of periodically driven systems with time-dependent Hamiltonians. They provide various routes to a non-trivial topological system in the condensed matter \cite{PhysRevLett.113.266801,PhysRevB.89.121401,Hubener2017}, photonics \cite{Kitagawa2012,Hafezi2013,Rechtsman2013,Mukherjee2017}, and cold atoms \cite{PhysRevLett.106.220402,Jotzu2014,PhysRevLett.122.253601,Guo:19}. Those dynamic systems can also be described by the topological invariants used in static systems \cite{Nathan_2015,PhysRevLett.114.056801,PhysRevX.6.021013,PhysRevLett.121.196401}, and obey bulk-edge correspondence \cite{PhysRevX.3.031005}. Beyond these, there are many unique characteristics due to the periodic driving, like time crystals \cite{PhysRevLett.117.090402,PhysRevLett.118.030401} and anomalous edge states \cite{PhysRevB.82.235114}.

Ultracold atoms in the optical lattice provide a controllable platform to simulate condensed matter \cite{PhysRevLett.112.086401,PhysRevA.95.033629,Songeaao4748,Jin_2019,PhysRevLett.121.265301}. An important breakthrough is the implementation of artificial gauge fields in the ultracold atoms system, which makes it possible to simulate topological phenomena, for instance, Weyl semimetal \cite{PhysRevLett.114.225301,Noh2017,Song2019}, Chern number in high-dimensional space \cite{Lohse2018} and Chiral edge states \cite{PhysRevA.85.063614,Leder2016,PhysRevLett.112.043001,Mancini1510}. As a method of achieving artificial gauge fields, the shaken optical lattice opens a path to study chiral edge states in the Floquet system \cite{PhysRevA.89.063628,arXiv:2002.09840}, with the advantages of simplicity, high efficiency and high controllability \cite{PhysRevLett.109.145301,PhysRevA.85.023637,PhysRevA.89.051605}.

In this paper, we demonstrate the simulation of edge states in position shaken finite-size honeycomb optical lattice. We periodically shake the position of the finite-size honeycomb optical lattice, which breaks the time-reversal-symmetry and causes topological non-trivial phases to appear in the system. Through adjusting the shaking frequency and shaking direction, the topological phase will transit. We find a pair of chiral edge states on different sides of lattice in the topological non-trivial area. Furthermore, at phase transition points we observe a pair of gapless boundary excitations with two chiral edge states on the same side of the lattice. Combining with the advantages of ultracold atoms in optical lattices, it might be a new approach to study chiral edge states in Floquet systems.

The remainder of this paper is organized as follows. In Sec.II, we introduce the model of the shaken honeycomb optical lattice and derive effective Hamiltonian of the shaken lattice. In Sec.III, we calculate the Chern number of the infinite-size lattice. In Sec.IV, we restrict the number of lattice sites and observe the chiral edge states in the finite-size lattice. Finally, we give a conclusion in Sec.V.

\begin{figure*}[htp]
	\includegraphics[width=0.95\textwidth]{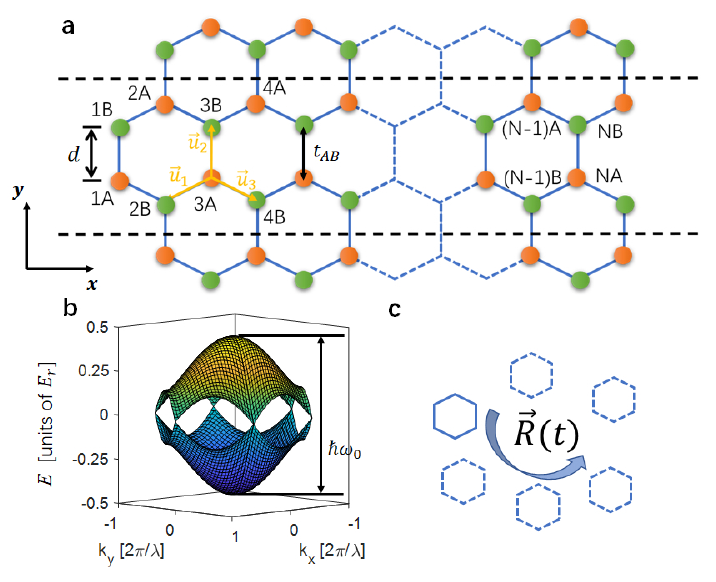}
	\caption{\textbf{Schematic diagram of the 2D shaken honeycomb lattice.} (a) The finite-size honeycomb optical lattice in our system. The orange circle represents the point A, and the green circle is point B. The lattice sites between the two black dashed lines form a repeating unit, which satisfies the periodic boundary condition marked by the black dashed lines in the $y$ direction, and has $N$ sites in $x$ direction. $t_{AB}$ denote the nearest-neighbor tunneling coefficient, and $d$ is the distance between nearest-neighbor points. $\vec{u}_1,\vec{u}_2,\vec{u}_3$ are vectors between nearest-neighbor points.  (b) The energy spectrum of the honeycomb optical lattice's lowest two bands. Parameters: $V_{A}=V_{B}=13.0 ~Er, d=\frac{2\sqrt{3}}{9}\lambda=409.53 ~{\rm nm}$, where $\lambda=1064 ~{\rm nm}$ is the wavelength of laser beams, and $V_{A}$ and $V_{B}$ are the potential depth of point A and point B respectively. $Er$ is photon recoil energy. $\hbar\omega_0$ represents the width of the lowest two bands. (c) Schematic diagram of the position shaking trajectory $\vec{R}(t)$. The blue arrow indicates the form of shaking. The solid blue line is initial position of the lattice cell, and the blue dashed line is the position of the cell during shaking.
	}\label{fig:fig1}
\end{figure*}

\section{Model description and Effective Hamiltonian}
\subsection{Model description}

We constructed a model of finite-size honeycomb optical lattice, as Fig. \ref{fig:fig1}(a) shows. In the honeycomb lattice, there are two kinds points: the orange circle represents the point A, and the green circle is point B. $d=\frac{2\sqrt{3}}{9}\lambda$ is the distance between nearest-neighbor points, where $\lambda$ is the wavelength of laser beams forming the honeycomb optical lattice. 
$N$ is the number of lattice sites in $x$ direction. A periodic boundary condition is used in the $y$ direction, marked by the black dashed lines. 
Between the two black lines, there are $2N$ sites, and these sites form a repeating unit of the lattice. The properties of the honeycomb lattice can be reflected by a repeating unit.

Neglecting the weak atomic interactions, the single-particle two-bands tight-binding Hamiltonian of the 2N sites in a repeating unit can be written:  
\begin{eqnarray}
\label{H 0}
\hat{H_0}=\sum_{i}V_i c_i^\dag c_i +\sum_{i\ne j}t_{ij} c_i^\dag c_j.
\end{eqnarray}
The $i$ ($j$) marks the site in a repeating unit, represented by an array $(C,n)$, where $C$=A,B and $n=1,2,3,...N$. 
$\sum_{i\ne j}$ denotes summation over pairs of different sites.
$V_i$ is the potential depth at site $i$. 
$t_{ij}$ denotes the tunneling coefficient between site $i$ and $j$.  
The operators $c_{i}^\dagger$ and $c_{i}$ denote the creation and annihilation operator at site $i$, and so is $c_{j}^\dagger$ and $c_{j}$.

Fig. \ref{fig:fig1}(b) shows the lowest two bands of the honeycomb lattice, which is described by Eq. (\ref{H 0}) and $N$ tends to infinity. In the following calculation, the potential depths of point A and B are both $13.0 ~Er$, and the nearest-neighbor tunneling coefficient $t_{AB}=0.15 ~Er$, which can be calculated by the overlapping integral of the wannier function \cite{PhysRevB.56.12847}. $Er=\frac{(2\pi \hbar)^2}{2M\lambda^2}$ is photon recoil energy, and $\lambda=1064 ~{\rm nm}$. $M$ is the mass of the $^{87}$Rb atom. In the figure, 
these two bands are the form of the trigonometric function, and the gap is closed. So in the origin of momentum space (the $x,y$ component of momentum $k_x=0$, $k_y=0$), the distance between the two bands is the largest. 
We define the largest distance as the width of the two bands as $\hbar\omega_0$. 

If we shake the optical lattice along track $\vec{R}(t)$, in optical lattice reference system, the atoms receive an inertial force $\vec{F}(t)=-M\ddot{\vec{R}}(t)$. So the Hamiltonian of the shaking honeycomb optical lattice is written as:
\begin{eqnarray}
\label{H}
\hat{H}=\hat{H_0}+\hat{H_1}(t),
\end{eqnarray}
where
\begin{eqnarray}
\label{H1}
\hat{H_1}(t)=-\sum_{i}\vec{F}(t)\cdot \vec{r}(i) c_i^\dag c_i.
\end{eqnarray}
$\vec{r}(i)$ is the position vector of site $i$.

We choose the shaking trajectory $\vec{R}(t)$ as an ellipse $(A_1cos\omega t,A_2cos(\omega t+\psi))$, shown in Fig. \ref{fig:fig1}(c), where $A_1cos\omega t,A_2cos(\omega t+\psi)$ are $x,y$ component of $\vec{R}(t)$. $\omega$ is the shaking frequency and $A_1,A_2$ are the shaking amplitudes. Shaking phase $\psi$ and shaking amplitude ratio $A_1/A_2$ jointly decide the direction of shaking. This shaking introduces an  artificial gauge field in the Floquet system, which is equivalent with the electromagnetic field in condensed matter systems.

The above 2D shaken optical lattice can be constructed with three linearly polarized laser beams, which are  perpendicular to lattice plane, with an enclosing angle of 120$^\circ$ to each other, which has been demonstrated in recent works \cite{Flschner1091,jin2019dynamical}. The total potential energy of optical lattice is written as
\begin{eqnarray}
\label{V_lattice}
V(\vec{r})=
V_{\rm 0}\sum_{i',j'}{\rm cos}\left[(\vec{k_{i'}}-\vec{k_{j'}})\cdot \vec{r}-(\theta_{i'}-\theta_{j'})\right].
\end{eqnarray}
where $i',j' =1,2,3$ represent three directions of wave vectors of laser, and $\vec{k_1}= (\sqrt{3}\pi,-\pi)/\lambda$, $\vec{k_2}= (-\sqrt{3}\pi,-\pi)/\lambda$, $\vec{k_3}= (0,2\pi)/\lambda$ are wavevectors. $\lambda$ is the wavelength of the laser beam, $V_{\rm 0}$ denote the potential energy of the lattice. The three angles $\theta_1,\theta_2,\theta_3$ represent the relative phases of the three laser beams. The shaking can be achieved by periodically modulating the three angles  with time as $(\theta_{i'}-\theta_{j'})=(\vec{k_{i'}}-\vec{k_{j'}}) \cdot \vec{R}(t)$, and the potential energy of optical lattice is rewritten as \cite{Guo19,zhou2020}:
\begin{eqnarray}
\label{V_lattice_t}
V(\vec{r})=
V_{\rm 0}\sum_{i',j'}{\rm cos}\left[(\vec{k_{i'}}-\vec{k_{j'}})\cdot (\vec{r}-\vec{R}(t))\right].
\end{eqnarray}
This potential energy means the lattice shake along $\vec{R}(t)$.

The size of the optical lattice is determined by the waist of the Gaussian beam, and the number of lattice sites in the light waist area can be considered as $N$. Generally, there are hundreds of sites in an optical lattice in our past experiment \cite{Guo19}.

\subsection{Effective Hamiltonian}
Above, we describe the model of the shaken finite-size optical lattice and give the Hamiltonian $\hat{H}$ with time. In this section, for further calculation, we introduce the method to calculate time-averaged effective Hamiltonian $H_{eff}$ using Floquet theory.

First, we use an unitary transformation to change the $\hat{H}$ into the Hamiltonian in laboratory reference system $\hat{H'}$ as:
\begin{eqnarray}
\label{H'}
&&\hat{H'}=\hat{U}\left(\hat{H}(t)-\mathbbm{i} \hbar\dfrac{\partial}{\partial t}\right)\hat{U}^\dag-\left(-\mathbbm{i} \hbar\dfrac{\partial}{\partial t}\right) \nn \\ 
&&=\sum_{i}V_i c_i^\dag c_i-\sum_{i\ne j} t_{ij} \exp\left(\mathbbm{i} z_{ij} sin(\omega t+\phi_{ij}) \right)c_i^\dag c_j.
\end{eqnarray}
$\hat{U}$ is a unitary operator:
\begin{eqnarray}
\label{U}
\hat{U}=\exp\left(\int_{0}^{t}d\tau\dfrac{\mathbbm{i}}{\hbar} \sum_{i}\vec{F}(\tau)\cdot \vec{r}(i) c_i^\dag c_i\right).
\end{eqnarray}
$z_{ij}$, $\phi_{ij}$, $\vec{r}_{ij}$, $\rho_{ij}$ are defined as:
\begin{eqnarray}
\label{zij}
z_{ij}=-M\omega\cdot\rho_{ij}/\hbar,
\end{eqnarray}
\begin{eqnarray}
\label{phi}
\phi_{ij}=arcsin{[\vec{r}_{ij}\cdot(0,A_2sin\psi)/\rho_{ij}]},
\end{eqnarray}
\begin{eqnarray}
\label{rrij}
\vec{r}_{ij}=\vec{r}(i)-\vec{r}(j),
\end{eqnarray}
\begin{eqnarray}
\label{rho}
\rho_{ij}=\sqrt{[\vec{r}_{ij}\cdot(A_1,A_2cos\psi)]^2+[\vec{r}_{ij}\cdot(0,A_2sin\psi)]^2}.
\end{eqnarray}
The symbol $(A_1,A_2cos\psi)$ represents a 2D vector.

Next, using a Jacobi-Anger expansion: $e^{\mathbbm{i}z\sin\theta}=\sum_{n=-\infty}^{+\infty}\mathcal{J}_n(z)e^{\mathbbm{i}n\theta}$ where $J_n$ represents  $n$th order Bessel function of the first kind, we can rewrite $H'$ as an $e$-exponent form:

\begin{eqnarray}
\label{H2'}
\hat{H'}=
\sum_{i}V_i c_i^\dag c_i-&&\sum_{n=-\infty}^{+\infty}\sum_{i\ne j} t_{ij}\mathcal{J}_n(z_{ij}) \exp\left(\mathbbm{i} n (\omega t+\phi_{ij}) \right)c_i^\dag c_j \nn \\
&&=H_{f0}+\sum_{n\ne 0}H_n \exp\left(\mathbbm{i} n (\omega t) \right),
\end{eqnarray}
where $H_{f0}=\sum_{i}V_i c_i^\dag c_i-\sum_{i\ne j} t_{ij}\mathcal{J}_0(z_{ij}) c_i^\dag c_j$, and $H_n=-\sum_{i\ne j} t_{ij}\mathcal{J}_n(z_{ij}) \exp\left(\mathbbm{i} n \phi_{ij} \right)c_i^\dag c_j$. $H_{f0}$ is constant term of $e$-exponential expansion, and $H_n$ is coefficient of $\exp\left(\mathbbm{i} n (\omega t) \right)$, where $n=\pm1, \pm2, \pm3 \cdots$.

Finally, we perform a high-frequency expansion on $H'$ to get the effective Hamiltonian $H_{eff}$ as:
\begin{eqnarray}
\label{H eff}
H_{eff}=H_{eff}^{(0)}+\dfrac{1}{\hbar\omega}H_{eff}^{(1)},
\end{eqnarray}
where
\begin{eqnarray}
\label{H eff001}
&&H_{eff}^{(0)}=H_{f0} \nn \\
&&=-\sum_{i}\sum_{\vec{u}_l} J_0(z_{\vec{u}_l})t_{AB}e^{-\mathbbm{i} \vec{k}\cdot \vec{u}_l}
c_k^\dagger (i)c_k (i+\vec{u}_l),
\end{eqnarray}
\begin{eqnarray}
\label{H eff002}
&&H_{eff}^{(1)}=\sum_{n=1}^{\infty} \frac{[H_{n},H_{-n}]}{n} \nn \\
&&=2\mathbbm{i}\sum_{i}\sum_{\vec{u}_l}\sum_{\vec{u}'_l} J_1(z_{\vec{u}_l})J_1(z_{\vec{u}'_l})t^2_{AB}e^{-\mathbbm{i} \vec{k}\cdot (\vec{u}_l+\vec{u}'_l)} \nn \\
&&\cdot sin(\phi_{\vec{u}_l}-\phi_{\vec{u}'_l})c_k^\dagger (i)c_k (i+\vec{u}_l+\vec{u}'_l). 
\end{eqnarray}
$\vec{u}_l$ ($\vec{u}'_l$) are vectors between nearest-neighbor points: $\vec{u_1}=(-\frac{\sqrt{3}}{2}d,-\frac{1}{2}d),\vec{u_2}=(0,d)$ and $\vec{u_3}=(\frac{\sqrt{3}}{2}d,-\frac{1}{2}d)$. The operators $c_k(i)$ denote the annihilation operators at point $(i,n)$ in momentum space. $c_k(i)=a_k(n)$ at point A or $c_k(i)=b_k(n)$ at point B.
In the calculation of Eq. (\ref{H eff001}) and (\ref{H eff002}), we keep up to nearest-neighbor terms and choose $A_1=A_2=d$ (More deltails see in Appendix A).

It is worth noting that the effective Hamiltonian Eq.(\ref{H eff}) is applicable to any value of $N$. On the one hand, if $N$ is infinite, the system has translational symmetry, and in momentum space each creation operator at point A is equivalent, and so is point B. Therefore, the effective Hamiltonian is $2\times2$ dimensional. On the other hand, when N is finite, the translation symmetry in the $x$ direction is broken.  
Hence in the lattice point A is not equivalent to each other, and so is point B. In other words, these 2N operators $c_i$ are not equivalent to each other, which causes a $2N\times2N$ dimensional Hamiltonian. (More details about effective Hamiltonian of the infinite-size system see in Appendix A)

\section{Chern number of infinite-size system}
We have derived the effective Hamiltonian of the shaken lattice. Next we study its topological properties. The topological properties of a system can be divided into properties of bulk states and properties of edge states, which are different but related. In this section, we study the bulk states' topological properties in the Floquet system, through calculating its Chern number under different shaking parameters.

The Chern number reflects topological properties of bulk states. And the number of lattice sites $N$ should be large to reflect the bulk's properties. Hence, for convenience, we choose $N \to \infty$. The system in infinite situation is equavalent to a two-band system, so we can rewrite the form of effective Hamiltonian (\ref{H eff}) as \cite{rspa.1984.0023,RevModPhys.82.3045}:
\begin{eqnarray}
\label{H eff2}
H_{eff,N \to \infty}=h_0\cdot \hat{I}+\vec{h} \cdot \hat{\vec{\sigma}},
\end{eqnarray}
where $\hat{I}= \begin{pmatrix} 1 & 0 \\ 0 & 1\end{pmatrix}$ is the identity matrix and $\hat{\vec{\sigma}}$ is Pauli matrix. $h_0$ is a scalar to describe this Hamiltonian, and $\vec{h}$ is the Bloch vector of the lower band. Therefore the Berry curvature of the system can be written as \cite{Tarnowski2019}:
\begin{eqnarray}
\label{berry curvature}
\vec{\Omega}(\vec{k})=\dfrac{1}{2} (\partial _{k_x} \hat{h}\times\partial _{k_y} \hat{h}) \cdot \hat{h},
\end{eqnarray}
where $\hat{h}=\vec{h}/|\vec{h}|$. The Berry curvature only has $z$ component, and $x,y$ component of the system are always zero. Through surface integral of Berry curvature in the first Brillouin zone, we can get Chern number of the system as:
\begin{eqnarray}
\label{chern number}
C=\int _S {\rm d}\vec{S} \cdot \vec{\Omega}(\vec{k}).
\end{eqnarray}
Here $S$ is the integration area, which is chosen as the first Brillouin zone. The direction of $\vec{S}$ is $k_z$ direction perpendicular to the $k_x-k_y$ plane.

\begin{figure}
	\includegraphics[width=0.5\textwidth]{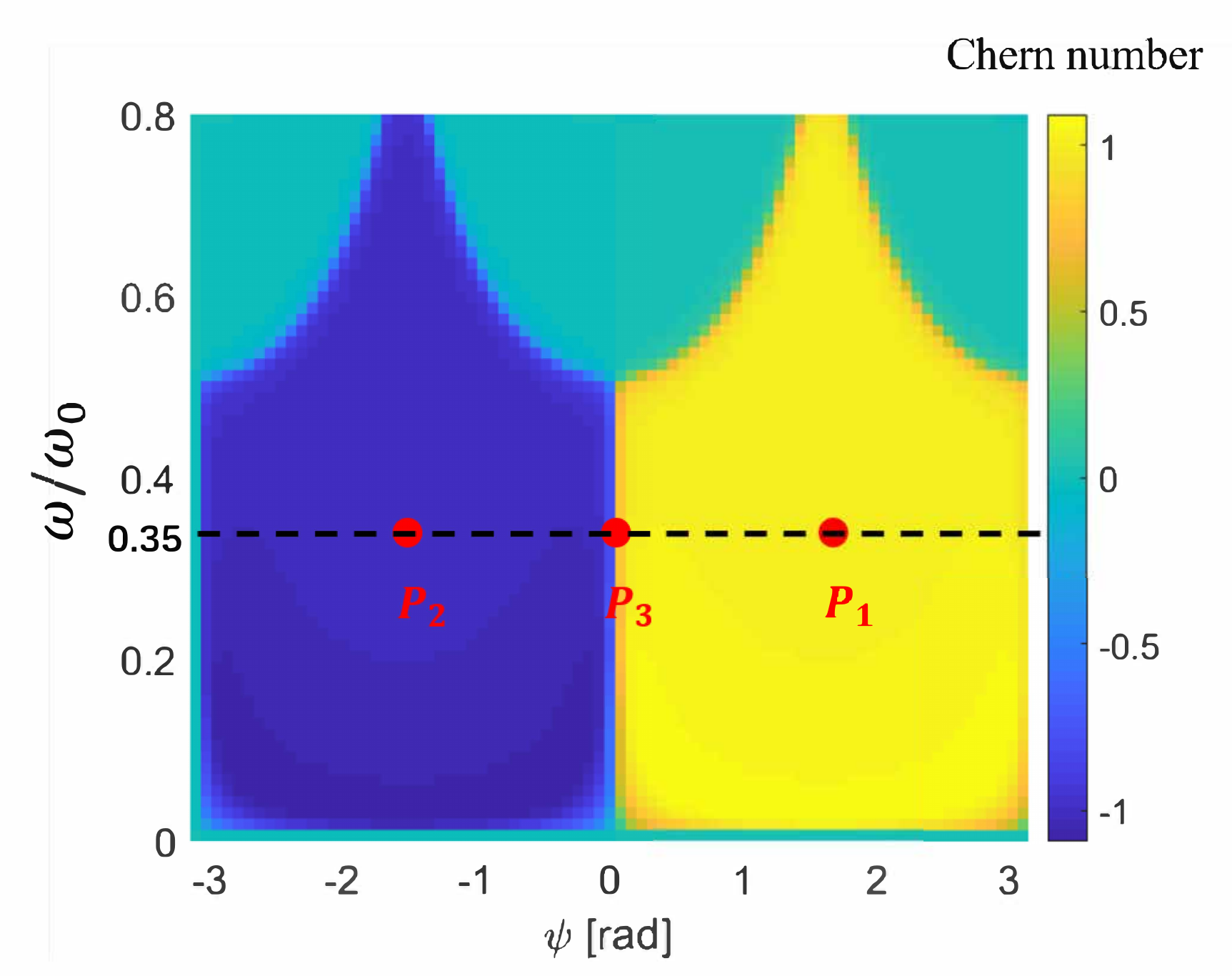}
	\caption{\textbf{The Chern number phase diagram.} The blue and yellow area represent topological non-trivial phases with Chern number $C=$ -1, +1, and green area represents topological trivial phase with $C=$ 0. The black dashed line marks the $\omega=0.35 \omega_0$, and red points $P_1$,$P_2$ and $P_3$ mark three situations of the phase diagram, which will be discussed below.
	}\label{fig:fig3}
\end{figure}

Removing non-nearest-neighbor tunneling coefficients and changing the shaking parameters $\omega$ and $\psi$, we can calculate the Chern number under different $\psi$ and $\omega$, shown in Fig. \ref{fig:fig3}. The figure shows the Chern number below $0.8 \omega_0$, and the yellow area and the blue area represent $C=+1$ and $C=-1$, respectively. Further, when $\psi$ is the opposite, the Chern number is also the opposite. The reason we focus on the area below $0.8 \omega_0$ is that when the shaking frequency $\omega$ is near $\omega_0$, the lowest two bands between point A and point B will be coupled with each other and the topological phase will become very complex.

\begin{figure}
	\includegraphics[width=0.4\textwidth]{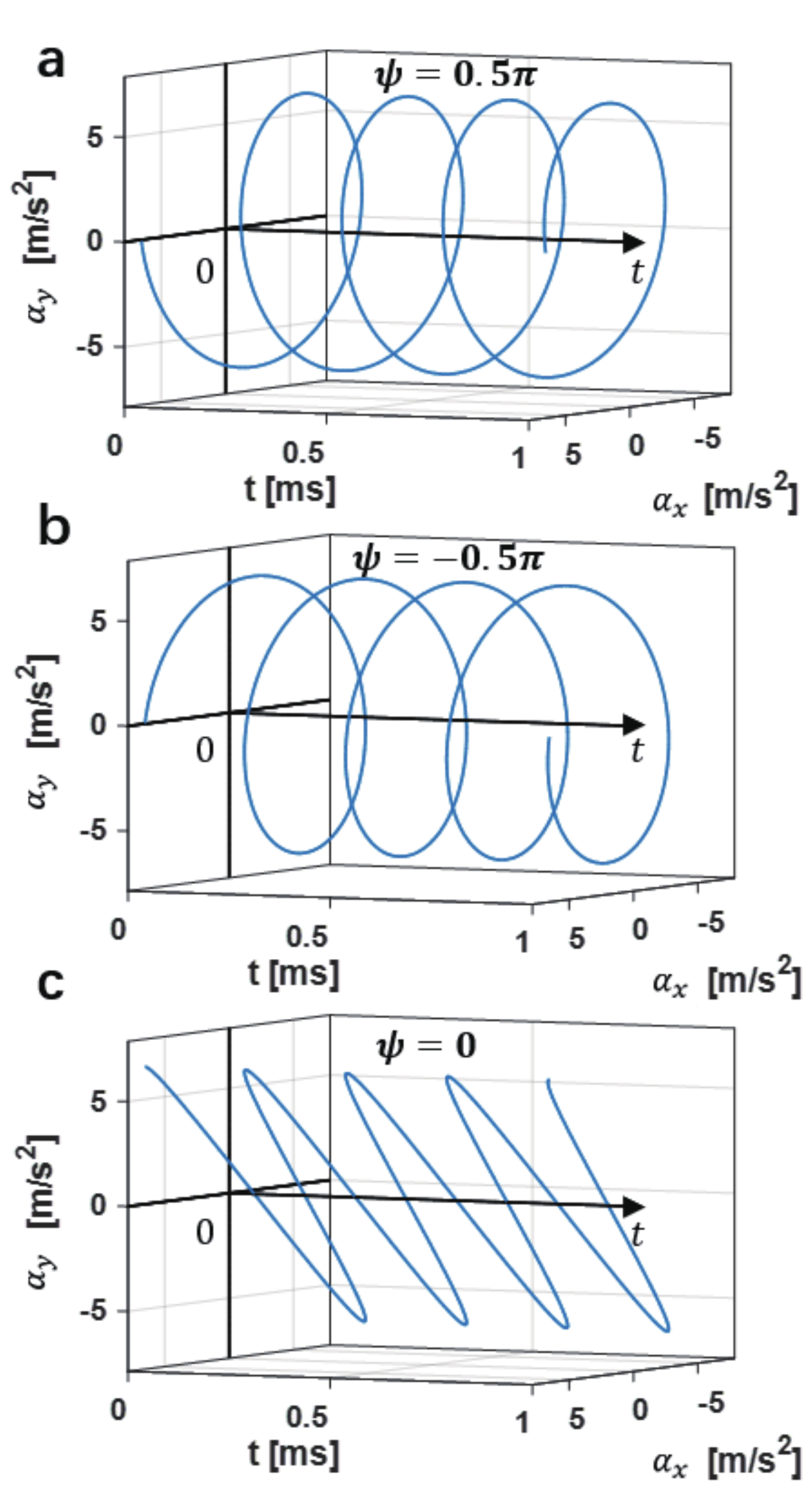}
	\caption{\textbf{Schematic diagram of equivalent electromagnetic acceleration over time.} $\alpha_x,\alpha_y$ is the x,y component of effective magnetic field acceleration $\vec{\alpha}$. The blue lines are the trajectories of equivalent electromagnetic acceleration over time. The parameters of figure (a), (b) and (c) are selected as point $P_1$, $P_2$ and $P_3$ in Fig. \ref{fig:fig3}, respectively.
	}\label{fig:fig2}
\end{figure}

In Fig. \ref{fig:fig3}, there are two kinds of area worth studying. The first is the topological non-trivial area, like point $P_1$ and $P_2$ where $\psi=\pm \pi/2$ and $\omega =0.35 \omega_0$. When $\pi>\psi >0$ (point $P_1$) and $-\pi<\psi <0$ (point $P_2$), Chern number is $\pm1$, respectively. And the other one is the phase transition points, taking point $P_3$ as an example, where $\psi=0$, $\omega =0.35 \omega_0$, and its Chern number can be calculated accurately as $0$. Further, we choose a typical line $\omega =0.35 \omega_0$, which is in the middle of the topological non-trivial area, and use the points on the line to describe the topological phase.

The shaking produces equivalent electromagnetic fields, which cause the non-trivial topological phases. We define the effective magnetic field acceleration to describe equivalent electromagnetic fields as:
\begin{eqnarray}
\label{acceleration}
\vec{\alpha}=(\alpha_x,\alpha_y)=\vec{F}(t)/M.
\end{eqnarray}

Fig. \ref{fig:fig2} shows $\vec{\alpha}$ at different $\psi$ over time. The Fig. \ref{fig:fig2}(a)(b) shows the situations at point $P_1$ and $P_2$ in Fig. \ref{fig:fig3}, and the paths of $\vec{\alpha}$ are circles in different directions. In Fig. \ref{fig:fig2}(c), which shows the situation at point $P_3$ in Fig. \ref{fig:fig3}, the $\vec{\alpha}$ oscillates along a straight line with time increasing. When $\psi \ne 0,\pm0.5\pi$, $\vec{\alpha}$ changes along an elliptical path with time. The straight path is equivalent to an oscillating electric field, and the system satisfies time reversal symmetry. While the circular or elliptical path break time reversal symmetry, the effect of which is equivalent to a magnetic field.

\begin{figure}
	\includegraphics[width=0.5\textwidth]{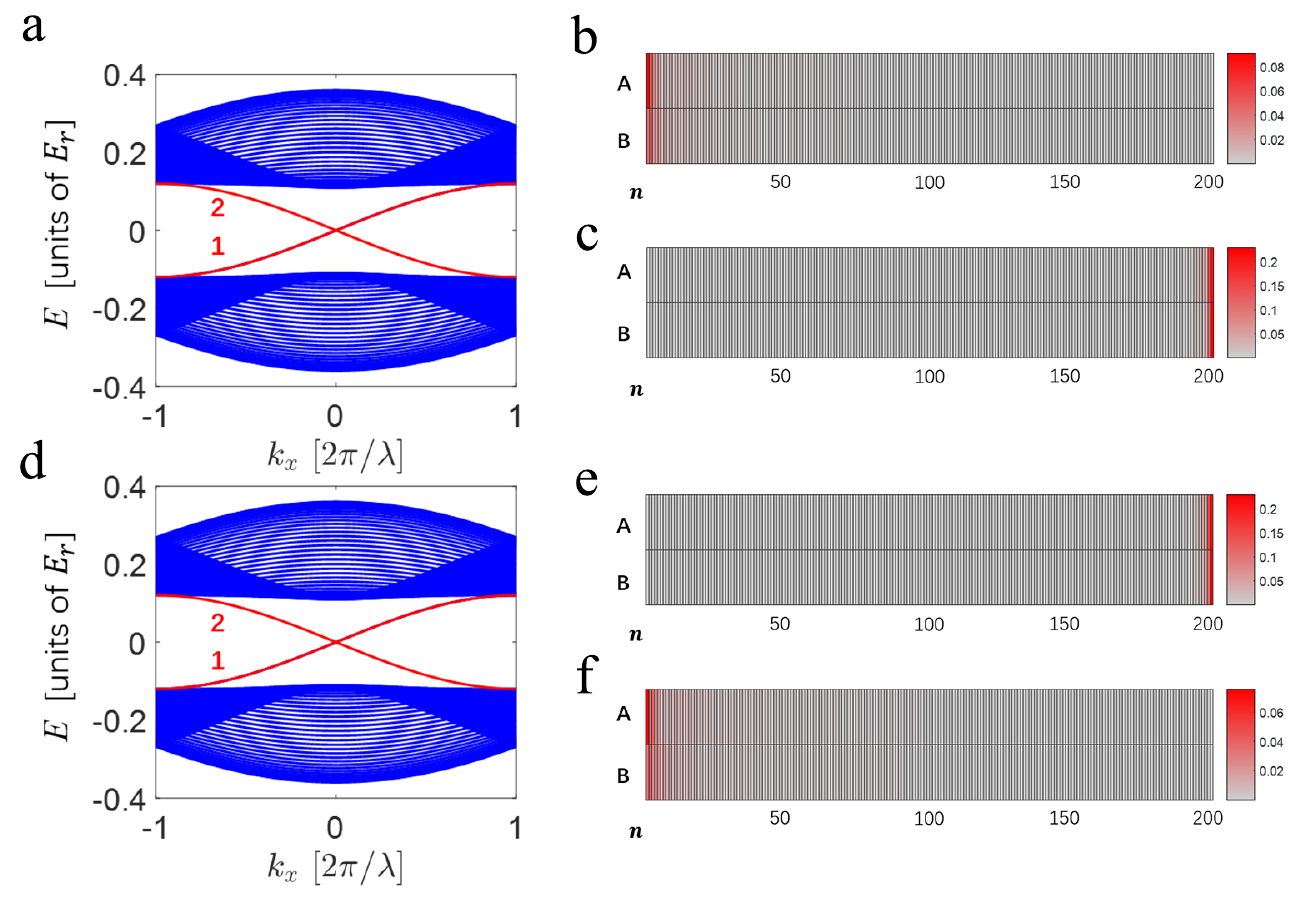}
	\caption{\textbf{The band structure and atomic density distribution of point P$_1$ and P$_2$ in Fig. \ref{fig:fig3}.} (a) The band structure of system at $k_y=0$ plane at point $P_1$. The number of point N=200. Blue lines are bulk bands, and the two red bands 1 and 2 are chiral edge states. There are a total of 400 bands in the system. Due to the limited picture size, only 100 bulk bands and 2 edge bands are drawn here. (b),(c) The atomic density distributions of band 1 and 2 at point $P_1$. The abscissa $n$ represents the number of the points, and the ordinate indicates that the point is A or B. (d-f) are the band structure and atomic density distributions corresponding to point $P_2$.
	}\label{fig:fig4}
\end{figure}

\section{Chiral edge states of finite-size system}
In the above section, we study the bulk states' topological properties of the system and find topological non-trivial phases and topological phase transition points. In the topological non-trivial area, it can be predicted, using classic bulk-edge correspondence, that there are chiral edge states. But at phase transition points, it is difficult to predict the situations of chiral edge state. In this section, we discuss the chiral edge states by band structure and atomic density distribution.

First, we study the edge states in the topological non-trivial area by band structure. We choose the number of points $N$ as $200$ in the following calculation, and solve the eigenvalue equation of the effective Hamiltonian $H_{eff}$ at point $P_1$ and $P_2$. Their band structures are shown in Fig. \ref{fig:fig4} (a) and (d). In these two figures, there are $2\times N$ bands. $2\times (N-1)$ of them can be divided into upper and lower blocks marked by blue lines, which correspond to the upper and lower band in Fig. \ref{fig:fig1}(b). And there are two bands connecting the upper and lower bulk bands, marked by red lines, and we call them band $1$ and $2$.

\begin{figure}
	\includegraphics[width=0.5\textwidth]{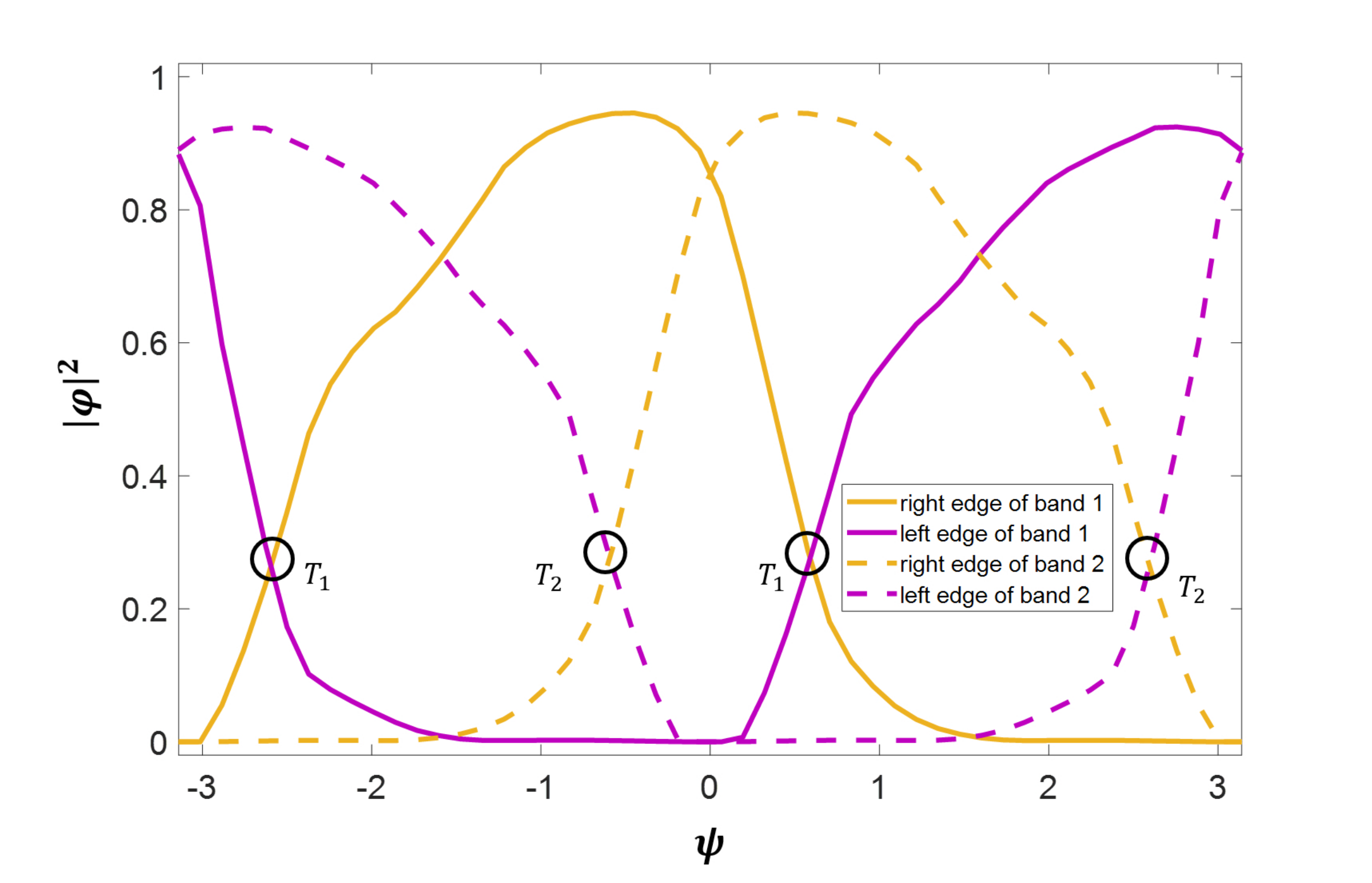}
	\caption{\textbf{The atomic density distributions of band 1 and 2 along $\bm{\omega=0.35 \omega_0}$ line.} Parameter: N=200, $\omega=0.35 \omega_0$. The lines are the fitted line of the calculated data points. The solid line and dashed line marks the band 1 and 2. The yellow and purple line represents the atoms in the six points on the right and left edge. The circle $T_1,T_2$ represent the transition point of edge band 1 and 2, respectively.
	}\label{fig:fig5}
\end{figure}

Next we show the atomic wave function location of band $1$ and $2$ in real space. The eigenstate $|c_i\rangle$ at site $i$ can be derived from Fourier transform of eigenstate in momentum space $|c_k(i)\rangle$:
\begin{eqnarray}
\label{eigenstate}
|c_i\rangle=\zeta(i)\sum_{k}|c_k(i)\rangle e^{-\mathbbm{i}\vec{k}\cdot \vec{r}(i)}.
\end{eqnarray}
$\zeta(i)$ is normalization coefficient. $|c_k(i)\rangle$ can be calculated by solving the eigenvalue equation of effective Hamiltonian (\ref{H eff}).

The eigenstates of two edge states at point $P_1$ and $P_2$ are shown in Fig. \ref{fig:fig4} (b-c) and (e-f). The abscissa is position $n$, and the ordinate represents point A or B. The colors in the figure represent the density of atomic distribution. The redder the block is, the higher the atomic density becomes. In Fig. \ref{fig:fig4} (c) and (e) there are more than $88\%$ atoms at the six points on the right edge, and the density is close to 0 away from the edge. While in Fig. \ref{fig:fig4} (b) and (f), there are also more than $50\%$ atoms at the twelve points on the right edge. It shows that band $1$ and $2$ are chiral edge states. Further, the edge states of band $1$ and $2$ are chirally symmetrical, because the optical lattice is chirally symmetrical. The ratio of atoms at the edge will increase with N.

In order to further study the chiral edge states in the topological non-trivial area, we calculate the atomic density distribution along the black dashed line $\omega=0.35 \omega_0$ in Fig. \ref{fig:fig3}. Shown in Fig. \ref{fig:fig5}, the yellow lines indicate the ratio of atoms on the left six points of the lattice to total atoms, and the purple lines are corresponding to the right edge. 
In the figure, at the points of highest atomic density, over 90\% atoms appear at one edge of the lattice. And at point P$_1$ and P$_2$, there are over 70\% atoms at the edge. This further explains the band $1$ and $2$ in Fig. \ref{fig:fig4} are edge bands.

\begin{figure}
	\includegraphics[width=0.45\textwidth]{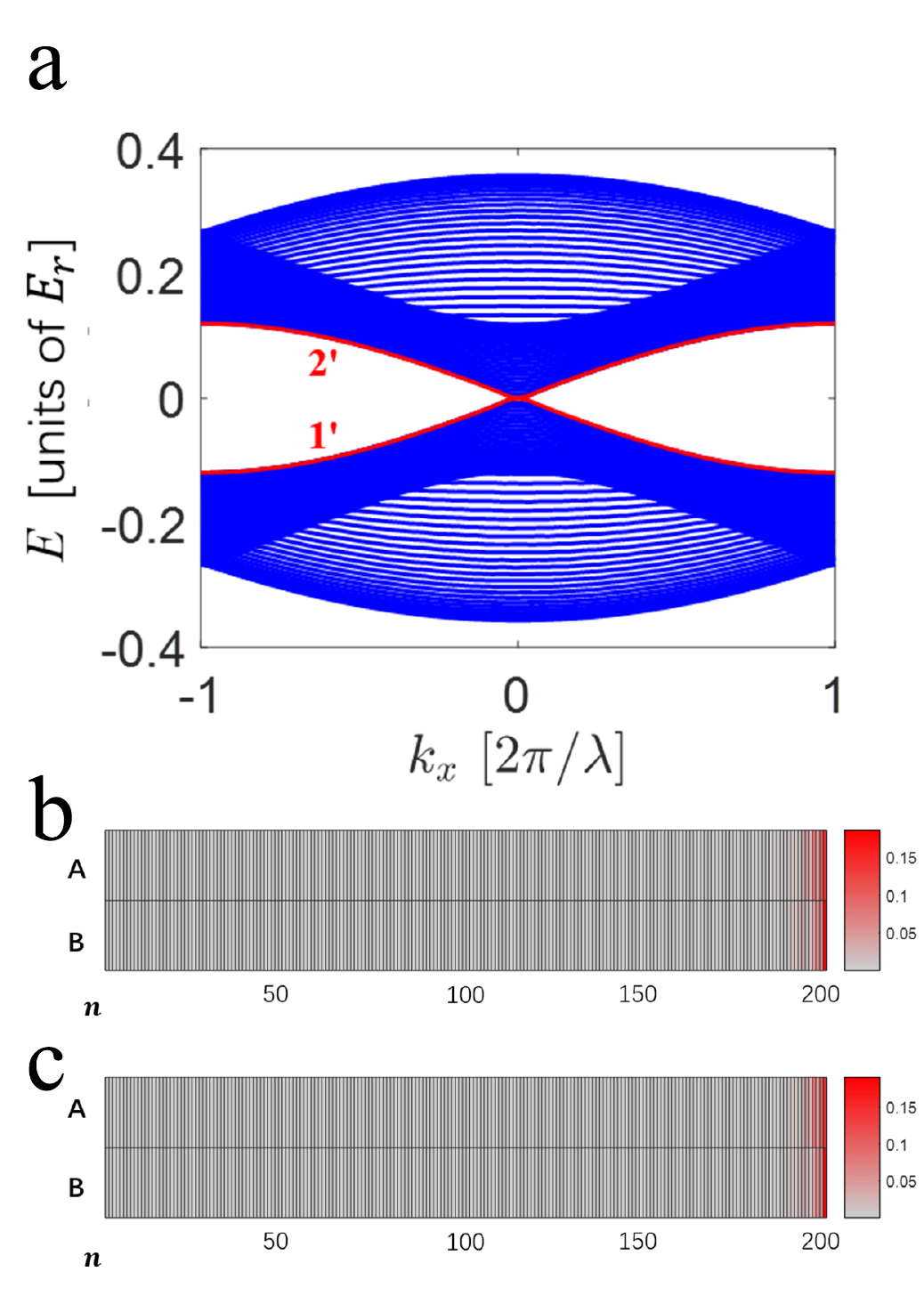}
	\caption{\textbf{The band structure and atomic density distribution of point P$_3$ in Fig. \ref{fig:fig3}.} (a) The band structure of system in $k_y=0$ plane at point $P_3$. The number of point N=200. Blue lines are bulk bands in the system, and the two red bands 1 and 2 are chiral gapless boundary excitations. Due to the limited picture size, only 100 bulk bands and 2 edge bands are drawn here. (b) (c) the atomic density distributions of band $1'$ and $2'$, at point P$_3$. The meaning of abscissa and ordinate is the same as Fig. \ref{fig:fig4}.
	}\label{fig:fig6}
\end{figure}

The transition points of bulk states, marked by Chern number appear at $\psi=0$. However, in Fig. \ref{fig:fig5}, the transition points of edge states, where atoms are distributed from left to right, are different from the transition points of bulk bands. The transition point $T_1$ and $T_2$ are marked by black circles. In band 1, when $\psi$ is from $-\pi$ to $\pi$, there are two phase transitions, and the difference of $\psi$ between the two $T_1$ points is $\pi$, so is $T_2$. Next, we focus on the two transition points around $\psi=0$. The $T_1$ point is to the left of the zero of $\psi$, and $T_2$ is to the right of the zero of $\psi$. Thus, around zero of $\psi$, the atoms of band 1 and 2 are distributed on the same side of the lattice.

\begin{figure}
	\includegraphics[width=0.45\textwidth]{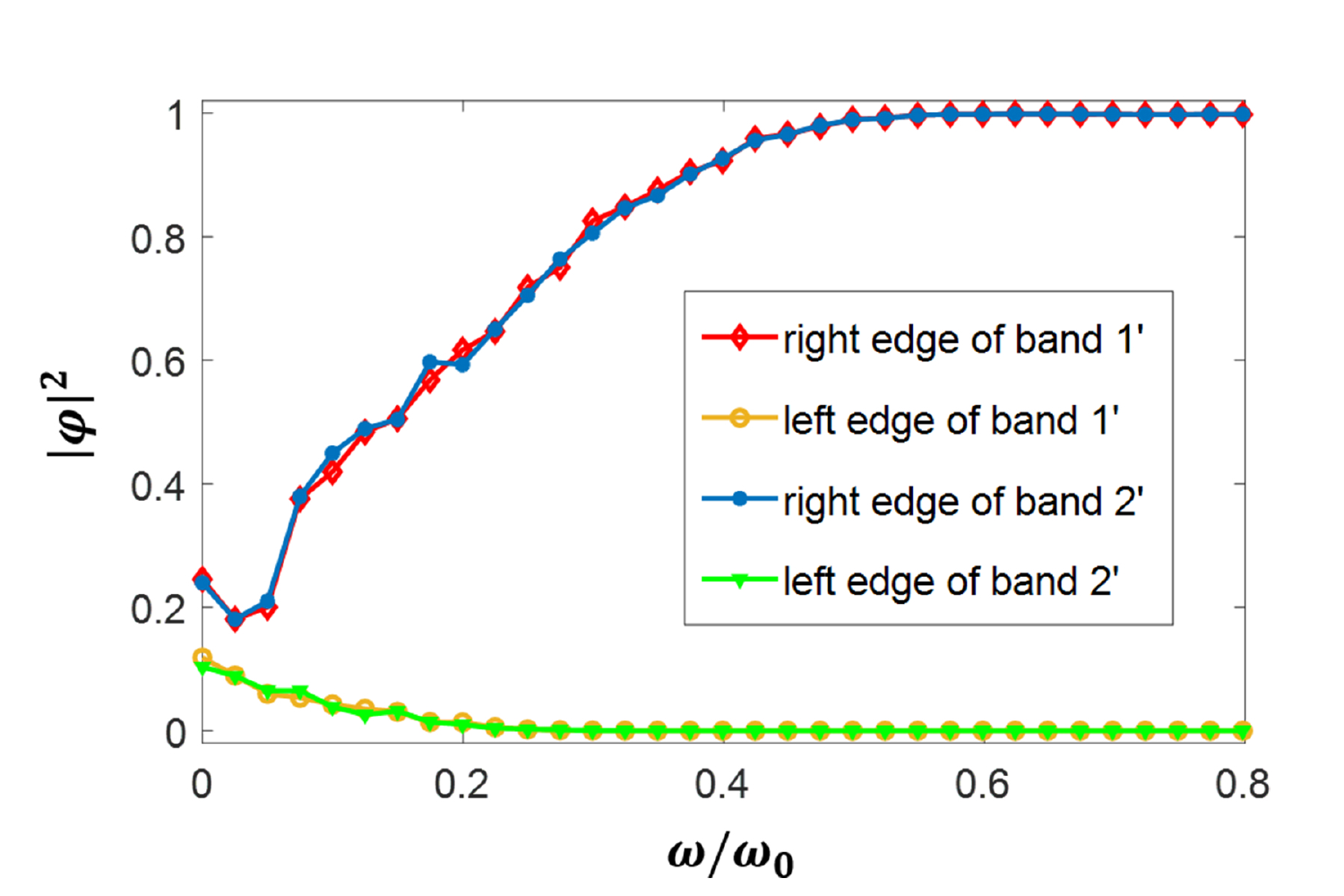}
	\caption{\textbf{The atomic density distributions of gapless boundary excitations along $\bm{\psi=0}$ line.} Parameter: N=200, $\psi=0$. The points are calculated data, and the solid lines are the fitted lines of these points. The red line and yellow line mark atomic ratio of band 1' in the six pionts on the left and right edge. The blue and green line corresponds to band 2'.
	}\label{fig:fig7}
\end{figure}

On the line $\psi=0$, taking point P$_3$ as an example, the band structure and atoms distribution of this point is shown in Fig. \ref{fig:fig6}. The bulk bands in Fig. \ref{fig:fig6}(a) are closed. Between the bulk bands there are two gapless excitations, marked by red lines, and we name them as band $1'$ and $2'$. These gapless excitations generally locate at the edges of the systems, because the quantum Hall states contain no bulk gapless excitations \cite{PhysRevB.25.2185,PhysRevB.41.12838}. We calculate the distribution of atoms to show the location of band $1'$ and $2'$. It is shown in Fig. \ref{fig:fig6} (b) and (c), corresponding to band $1'$ and $2'$, respectively. In the figures, over 82\% atoms are at six points on the right edge. It is consistent with our analysis of Fig. \ref{fig:fig5} that when $\psi=0$ the two edge bands will appear on the same edge. By the way, when $\psi=\pi$, the two chiral gapless boundary excitations will both appear at the left edge.
Consequently the two bands are indeed chiral gapless boundary excitations.

In order to fully describe the phase transition points of Chern number, we calculate the distribution of atoms along the line $\psi =0$, as Fig. \ref{fig:fig7} shows. In the figure, the yellow and green lines represent the atoms in the six points on the left edge of band $1'$ and $2'$, and the red and blue lines represent the atoms in the six points on the right edge of band $1'$ and $2'$. With shaking frequency $\omega$ increasing, the atoms gradually spread to the right edge of lattice, and finally over $99\%$ atoms are distributed on the right edge.

When $-\pi<\psi<\pi$, only at the transition points $\psi=0$, this system has time-reversal symmetry, which causes Chern number at transition points to be 0. And the potential depths of point A and B are the same, so the gapless boundary excitations will appear at transition points. The two gapless boundary excitations are on the same edge of the lattice, which means the chiral symmetry of this system is broken at these points. The reason of this broken chiral symmetry needs further research.

\section{Conclusion}
In summary, we propose a scheme to realize chiral edge states of Floquet system in position shaken honeycomb optical lattice. We shake the honeycomb optical lattice to get non-trivial topological phases, and observe edge states at non-trivial topological phases and phase transition points. Through deriving the effective Hamiltonian of the Floquet system, the Chern number under different shaking parameters of the system and corresponding chiral edge states are calculated. In the non-trivial topological area, we find a pair of chiral edge states at different sides of the optical lattice, and gapless boundary excitations at the same side of optical lattice around phase transition points. This Floquet system obeys the bulk-edge correspondence. However, the transition points of edge states and bulk states are inconsistent, which causes the gapless boundary excitations appearing on the same side. Our work might be a promising route to study edge states in Floquet system, and provides more detailed insight into the gapless boundary excitation.

\section{Acknowledgement}
We thank Xiaopeng Li and Wenjun Zhang for helpful discussion. This work is supported by the National Basic Research Program of  China (Grant No. 2016YFA0301501) and the National Natural Science Foundation of China (Grants No. 61727819, No. 11934002, No. 91736208, and No. 11920101004).

\appendix
\addcontentsline{toc}{section}{Appendices}\markboth{APPENDICES}{}
\begin{subappendices}
	
\section{Effective Hamiltonian of the finite-size system}
In the derivation of Eq.(\ref{H eff001}) and Eq.(\ref{H eff002}), it need to fourier transform $c_i^\dagger$ and $c_i$, the creation and annihilation operators in real space, to get $c_k(i)^\dagger$ and $c_k(n)$, the corresponding creation and annihilation operators in momentum space as:
\begin{eqnarray}
\label{FFFF}
c_i=\dfrac{1}{\sqrt{\Theta}}\sum_k c_k(i) e^{-\mathbbm{i}\vec{k}\cdot\vec{r}},
\end{eqnarray}
where $\Theta$ is the number of lattice sites. For finite-size system, the system doesn't have translational symmetry, and the operator every $c_k(i)$ is different. Within periodic boundary condition in y direction, there are 2N points (N point A and N point B). So the effective Hamiltonian has 2N eigenstates $|c_k(i)\rangle$. 

For example, if we choose $N=3$ and use a table to describe the effective Hamiltonian, the effective Hamiltonian can be written as:

\begin{minipage}{0.5\textwidth}
	\includegraphics[width=0.9\textwidth]{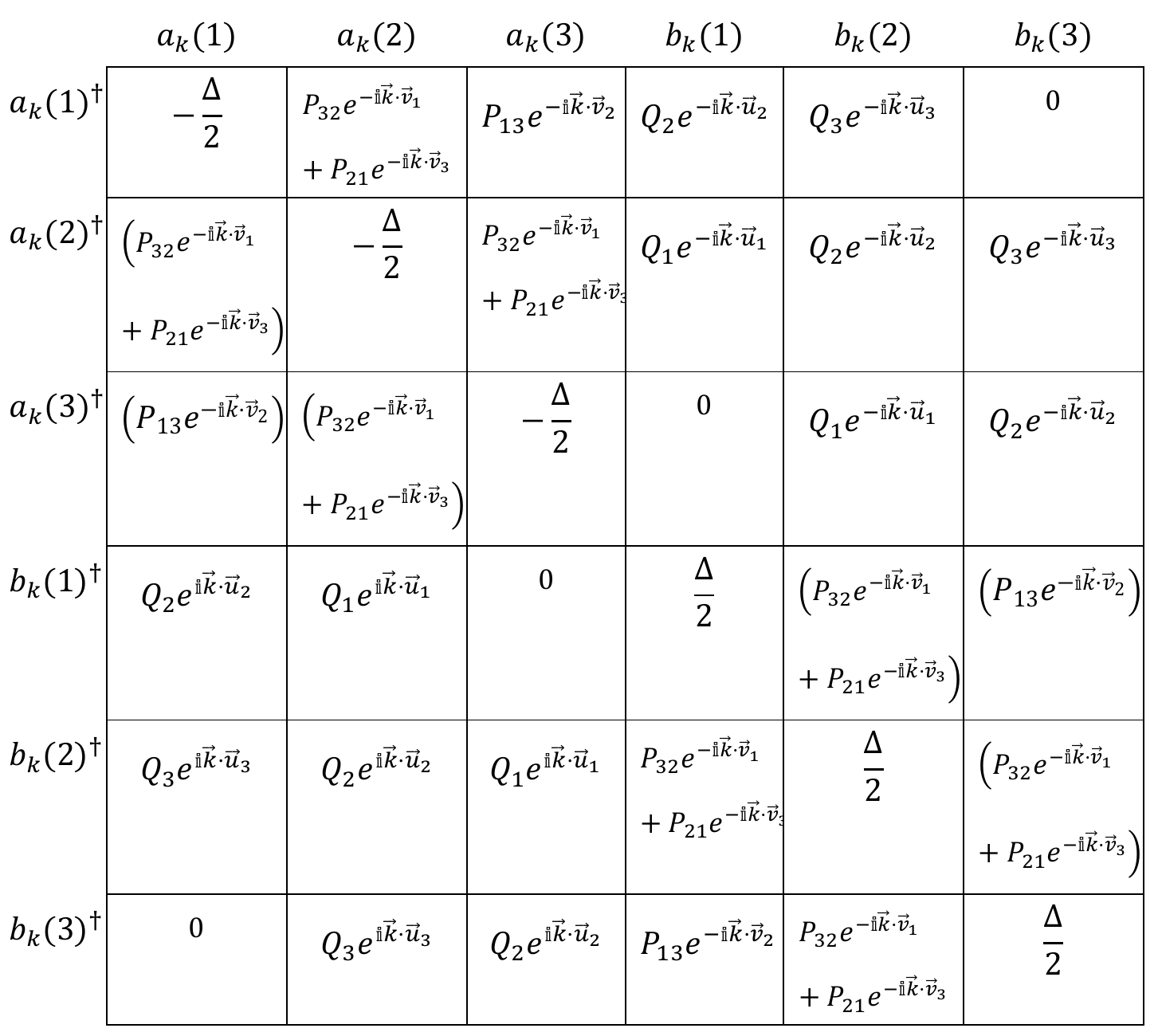}
\end{minipage}

\leftline{where}

\begin{eqnarray}\
\label{PPPPQQQ}
&&\vec{v_1}=\vec{u_3}-\vec{u_2}, 
\vec{v_2}=\vec{u_1}-\vec{u_3},
\vec{v_3}=\vec{u_2}-\vec{u_1},\nn \\
&&P_{32}=-(2i/\hbar \omega) J_1 (z_3 ) J_1 (z_2 ) t_3 t_2\nn \\
&&\cdot(A_2 u_{3y} sin(\psi) (-A_1\cdot u_{2x}-A_2\cdot u_{2y}\cdot cos(\psi))/(\rho_3 \rho_2  )\nn \\
&&+A_2 u_{2y} sin(\psi) \cdot (A_1 u_3x+A_2 u_{3y} cos(\psi))/( \rho_3 \rho_2)),\nn  \\
&&P_{21}=-(2i/\hbar \omega) J_1 (z_2 ) J_1 (z_1 ) t_2 t_1\nn  \\
&& \cdot(A_2\cdot u_{2y}\cdot sin(\psi)\cdot (-A_1 u_{1x}-A_2 u_{1y} cos(\psi))/(\rho_2 \rho_1  )\nn \\
&&+A_2\cdot u_{1y}\cdot sin(\psi)\cdot (A_1\cdot u_{2x}+A_2\cdot u_{2y}\cdot cos(\psi))/( \rho_2 \rho_1)),\nn  \\
&&P_{13}=-(2i/\hbar \omega) J_1 (z_1 ) J_1 (z_3 ) t_1 t_3\nn \\
&&\cdot (A_2\cdot u_{1y}\cdot sin(\psi)\cdot (-A_1\cdot u_3x-A_2\cdot u_{3y}\cdot cos(\psi))/(\rho_1 \rho_3  )\nn \\
&&+A_2\cdot u_{3y}\cdot sin(\psi)\cdot (A_1\cdot u_{1x}+A_2\cdot u_{1y}\cdot cos(\psi))/( \rho_1 \rho_3)),\nn \\
&&Q_1=-t_1 J_0 (z_1 ),
Q_2=-t_2 J_0 (z_2 ),
Q_3=-t_3 J_0 (z_3 ) .\nn
\end{eqnarray}

The $\Delta$ is the potential depth difference between point A and B, in the paper $\Delta=0$. $u_{lx},u_{ly}$ is the x,y component of $\vec{u_l}$, where $l=1,2,3$. $t_1=t_2=t_3=t_{AB}$ are the nearest-neighbor tunneling coefficients in direction $\vec{u_1}$, $\vec{u_2}$ and $\vec{u_3}$.

In the table, the abscissa represent annihilation operators, and the ordinate represent creation operators. The products of abscissa and ordinate represent the particle number operator. Each grid in the table indicates a term in the Hamiltonian, which is composed of the product of abscissa, ordinate and the data of it. For example, the data in the first row first column represent the form $-\frac{\Delta}{2}a_k(1)a_k(1)^\dagger$.

\section{Effective Hamiltonian of the infinite-size system}
For infinite-size system, the system has translational symmetry, so the operator $c_k(i)$ at every point A(B) is equivalent. And the effective Hamiltonian of system with finite size only has two eigenstates $|a_k\rangle$ and $|b_k\rangle$. Hamiltonian $H_{eff}^{(0)}$ and  $H_{eff}^{(1)}$ in Eq. (\ref{H eff001}) and (\ref{H eff002}) are simplified to:
\begin{eqnarray}
\label{H eff01}
&&H_{eff}^{(0)}=-\frac{\Delta}{2} \sum_{k}(a_k^\dagger a_k -b_k^\dagger b_k ) \nn \\
&&-\sum_{\vec{u}_l} J_0(z_{\vec{u}_l})t_{AB}e^{-\mathbbm{i} \vec{k}\cdot \vec{u}_l}
a_k^\dagger b_k +h.c. ,
\end{eqnarray}
\begin{eqnarray}
\label{H ef11}
&&H_{eff}^{(1)}=\frac{2\mathbbm{i}}{\hbar \omega}\sum_{\vec{u}_l}\sum_{\vec{u}'_l} J_1(z_{\vec{u}_l})J_1(z_{\vec{u}'_l})t^2_{AB}e^{-\mathbbm{i} \vec{k}\cdot (\vec{u}_l+\vec{u}'_l)} \nn \\
&&\cdot sin(\phi_{\vec{u}_l}-\phi_{\vec{u}'_l})c_k^\dagger (i)c_k (i+\vec{u}_l+\vec{u}'_l).
\end{eqnarray}
The effective Hamiltonian with two eigenstates can be rewritten as Eq.(\ref{H eff2}).

\end{subappendices}

\bibliographystyle{apsrev}
\bibliography{my}

\end{document}